\documentclass[11pt]{article}
\usepackage{graphicx,amsmath,bm, amsthm,mathrsfs,amssymb, braket, verbatim}
\usepackage[usenames]{color}
\usepackage{ulem,mathtools}
\usepackage{authblk}
\usepackage{cite}
\usepackage{tcolorbox}
\usepackage[section]{placeins}

\newcommand{\ee}{\end{equation}}

\newcommand{\reff}[1]{(\ref{#1})}
\newcommand{\beq}{\begin{equation}}
\newcommand{\eeq}[1]{\label{#1}\end{equation}}
\newcommand{\beqa}{\begin{eqnarray}}
\newcommand{\eea}{\end{eqnarray}}
\newcommand{\eeqa}[1]{\label{#1}\end{eqnarray}}
\newcommand{\beg}{\begin{equation*}}
\newcommand{\eeg}{\end{equation*}}

\newcommand{\bsplit}{\begin{split}}
\newcommand{\esplit}{\end{split}}

\usepackage{subfig} %To put many figures side by side
\usepackage{circuitikz} %To draw the electrical circuits
\usepackage[capposition=bottom]{floatrow} %To add notes under a figure
\usepackage{epigraph} %For quotes at the beginning of sections and text
\usepackage{appendix}
\usepackage{rotating}
\allowdisplaybreaks

\title{Wave packets in QFT: leading order width corrections to decay rates and clock behaviour under Lorentz boosts}

\author[]{Ariel Edery\thanks{aedery@ubishops.ca}}

\affil[]{Department of Physics and Astronomy, Bishop's University, 2600 College Street, Sherbrooke, Qu\'{e}bec, Canada, J1M 1Z7.\vspace{1em}}

\begin{document}
\date{}
\maketitle
\begin{abstract}
Decay rates in quantum field theory (QFT) are typically calculated assuming the particles are represented by momentum eigenstates (i.e. plane waves). However, strictly speaking, localized free particles should be represented by wave packets. This yields width corrections to the decay rate and to the clock behaviour under Lorentz boosts.  We calculate the decay rate of a particle of mass $M$ modeled as a Gaussian wavepacket of width $a$ and centered at zero momentum. We find the decay rate to be $\Gamma_0 \big[1- \frac{3 a^2}{4 M^2} +\mathcal{O}\big(\tfrac{a^4}{M^4}\big)\big]$ where $\Gamma_0$ is the decay rate of the particle at rest treated as a plane wave. The leading correction is then of order $\tfrac{a^2}{M^2}$. We then perform a Lorentz boost of velocity $v$ on the above Gaussian and find that its decay rate does not decrease \textit{exactly} by the Lorentz factor $\sqrt{1-v^2}$. There is a correction of order $\tfrac{a^2v^2}{M^2}$. Therefore, the decaying wave packet does not act exactly like a typical clock under Lorentz boosts and we refer to it is a ``WP clock" (wave packet clock). A WP clock does not move with a single velocity relative to an observer but has a spread in velocities (more specifically, a spread in momenta). Nonetheless, it is best viewed as a single clock as the wave packet represents a one-particle state in QFT. WP clocks do not violate Lorentz symmetry and are not based on new physics: they are a consequence of the combined requirements of special relativity, quantum mechanics and \textit{localized} free particles.         
\end{abstract}
%\thispagestyle{empty}
%\end{titlepage}
\setcounter{page}{1}
\newpage
\section{Introduction}\label{Intro}
Localized particles in Quantum Field Theory (QFT) are represented by wave packets\cite{Peskin,Burgess} and localized particles are what take part in decay and scattering processes\footnote{In this work we will focus on decay rates of wave packets and not scattering cross sections.}. In fact, replacing wave packets by momentum eigenstates (i.e. plane waves) leads to a well-known issue at a theoretical level: momentum eigenstates cannot be properly normalized in an infinite volume. This is usually handled using the ``finite-volume trick" \cite{Burgess} where one uses finite (but large) spatial volume $\Omega$ and interactions over a finite (but large) time interval $T$. One can then calculate the decay rate using momentum eigenstates because the divergences that naturally occur due to the squaring of delta functions are now regularized by $\Omega$ and $T$. In the end, this regularization dependence cancels and does not appear in quantities of interest. It is these calculations involving momentum eigenstates that are usually compared to decay experiments. This works in practice because the wave packets that appear in experiments have an energy width which is small compared to the particle decay width \cite{Burgess}. The detectors typically do not have the resolution to measure its effect. As pointed out in \cite{Peskin}, ``Real detectors have finite resolution...the measurement of the final-state momentum is not of such high quality that it can resolve the small variation of this momentum that results from the momentum spread of the initial wave packet". As stated in \cite{Burgess}, ``the wave packet is classical in the sense that the resolution of initial position and momentum measurements are much too large to push the limits of the uncertainty relations". Therefore, S-matrix calculations carried out in the plane wave limit have been usually adequate in practice for comparison with experiment. Nonetheless, we mention at the end of this introduction that the effects of wave packets may have already appeared in neutrino oscillation experiments. So the gap between experiment and wave packet analysis may already be showing signs of narrowing, at least for a certain type of experiment.     

In this work, we obtain analytical results for the leading order correction to decay rates due to the width of a Gaussian wave packet. There are at least three reasons to undertake such a calculation. The first reason is better theoretical understanding: it is beneficial to extract a leading order analytical result from an otherwise intractable integral that has only a numerical solution. The second reason is that the resolution of detectors and the design of experiments keeps improving so that the effects of wave packets on decay rates may become detectable in next generation particle decay experiments. The third reason might be the most important: we show that wave packets behave like a slightly different kind of clock under Lorentz boosts and we refer to it as a ``WP clock" (wave packet clock). This is interesting in and of itself and may spawn new concepts which we discuss in the conclusion. 

We first determine the leading order width correction to the decay rate of a particle at rest due to the width $a$ of a Gaussian wave packet centered at zero momentum. The correction is of order $\tfrac{a^2}{M^2}$ i.e. of order of the square of the ratio of the Compton wavelength to the inverse width of the Gaussian. We then perform a Lorentz boost of velocity $v$ on the original Gaussian wave packet and find that the decay rate does not decrease \textit{exactly} by the usual Lorentz factor $\sqrt{1-v^2}$. There is a small correction of order $\tfrac{a^2\,v^2}{M^2}$. This implies that the wave packet does not act like a typical clock under Lorentz boosts and we refer to it as a ``WP clock" (for wave packet clock). The distinctive feature of a WP clock is that it does not move with a single velocity relative to an observer but has a spread in velocities (more specifically, a spread in momenta). It should nonetheless be viewed as a single clock as the wave packet represents a one-particle state in QFT. It should be emphasized that there is no breakdown of Lorentz symmetry here. The WP clock does not stem from new physics but is an important consequence of combining the requirements of special relativity, quantum mechanics and \textit{localized} free particles.

We show that a Gaussian wave packet does not transform into a Gaussian under a Lorentz boost. We therefore calculate separately the decay rate of a Gaussian wave packet centered at velocity $v$ (i.e. momentum centered at $M\, \gamma\, v$ where $\gamma=(1-v^2)^{-1/2}$) and compare it to the decay rate of a Gaussian centered at zero momentum. Again, we find that the two decay rates are not related exactly by the Lorentz factor $\sqrt{1-v^2}$ except when the ratio of their respective widths have a particular value. 

In this work we do not consider corrections due to processes taking place over a finite-time interval (instead of an infinite time interval). This has been well studied in other work \cite{Tobita,Tobita2,Tobita3,Tajima}. More recently, in \cite{Ishikawa} a systematic study was undertaken to derive Fermi's golden rule in QFT within a  Gaussian wave packet formalism \cite{Shimomura,Araki1,Araki2,Yamamoto,Kasari}. They succeeded in separating bulk from boundary time contributions while maintaining the unitarity of the S-matrix in contrast to earlier work \cite{Stuck}. They showed that the bulk contribution leads to Fermi's golden rule while the boundary part can contribute deviations from it. This provided a solid framework in which to view the results of the aforementioned work\cite{Tobita,Tobita2,Tobita3,Tajima}. 

In \cite{Blasone} a wave packet analysis of quantum correlations in neutrino oscillations\cite{Giunti1,Giunti2, Giunti3} was undertaken to extend previous work \cite{Ming} that had been carried out using a plane wave analysis. The claim is that this wave packet analysis leads to a better agreement with the experimental results, most notably the MINOS experiments \cite{MINOS1,MINOS2} (the corrections due to the wave packet are more negligible for the Daya Bay experiments\cite{Daya1,Daya2}). In \cite{Cheng} neutrino wave packet effects for medium-baseline reactor neutrino oscillation (MBRO) experiments were also studied. Though it needs to be confirmed, effects of wave packets may already be showing up in experiments, in particular those involving neutrino oscillations. 

Our paper is organized as follows. We first determine the decay rate of a Gaussian wave packet centered at zero momentum and determine the correction to the plane wave result due to the width $a$ of the Gaussian. We then determine how a wave packet transforms under a Lorentz boost and calculate the decay rate of the Lorentz-boosted wave packet. This is where the WP clock is introduced. The Lorentz-boosted wave packet is not a Gaussian and we therefore calculate separately the decay rate of a Gaussian wave packet centered at velocity $v$ (i.e. momentum $M\,\gamma\,v$). The conclusion discusses an important and interesting problem that emerges from this work. The integrals we encounter in this paper do not have an exact analytical solution but can be evaluated via a power series expansion up to arbitrary accuracy. The details are relegated to Appendix A. 

\section{Decay rate of a wave packet}

The state $\ket{\phi}$ of a wave packet in quantum field theory can be expressed as a linear superposition of one-particle states \cite{Peskin}
\beq 
\ket{\phi}= \int \dfrac{d^3k}{(2\pi)^3} \dfrac{1}{\sqrt{2E_{\bf k}}}\phi({\bf k})\ket{\bf k}
\eeq{Packet}
where $\ket{\bf k}$ is a one-particle state of momentum ${\bf k}$, $E_{\bf k}$ is the relativistic energy $\sqrt{{\bf k}^2 +m^2}$ and $\phi({\bf k})$
is the Fourier transform of the spatial wavefunction. We use the Lorentz invariant normalization \cite{Peskin,Burgess} $\braket{{\bf p}|{\bf q}}=2E_{\bf p} (2 \pi)^3 \delta^3({\bf p}-{\bf q})$. We then obtain the conventional normalization $\braket{\phi|\phi}=1$ 
(where the sum of all probabilities is unity) if 
\beq
\int \dfrac{d^3k}{(2\pi)^3} |\phi({\bf k})|^2=1\,.
\eeq{Norm}
The differential decay rate of an unstable particle of mass $M$ treated as a plane wave (subscript ``0") with four-momentum $k$ (where $k^0=E_{\bf k}$) decaying into a set of final particles with four-momenta $p_f$ is given by \cite{Peskin}
\begin{align}
d\Gamma_0&=  \dfrac{1}{2 E_{\bf k}}\Big(\prod_{f} \dfrac{d^3{\bf p}_f}{(2 \pi)^3}\dfrac{1}{2 E_f}\Big)|\mathcal{M}(k\to\{p_f\})|^2(2 \pi)^4 \delta^4(k-\textstyle{\sum}\, p_f)\nonumber\\
&=\dfrac{1}{2 E_{\bf k}}\,d\xi
\label{DiffDecay}
\end{align} 
where 
\beq
d\xi=\Big(\prod_{f} \dfrac{d^3{\bf p}_f}{(2 \pi)^3}\dfrac{1}{2 E_f}\Big)|\mathcal{M}(k\to\{p_f\})|^2(2 \pi)^4 \delta^4(k-\textstyle{\sum}\, p_f)
\eeq{Invariant}
is Lorentz invariant and hence independent of ${\bf k}$ (e.g. if you evaluate it say at ${\bf k}=0$ then the result is valid for all ${\bf k}$). The dependence of the decay rate on ${\bf k}$ appears only in the prefactor $\frac{1}{2 E_{\bf k}}$. This factor is not Lorentz invariant and is the part responsible for time dilation i.e. that the lifetime for a particle increases with its speed. If we consider the unstable particle as a wave packet, we only need to integrate over this factor. The differential decay rate for the wave packet is then given by 
\beq
d\Gamma_{\text{packet}}= \Big(\int \dfrac{d^3k}{(2\pi)^3} |\phi({\bf k})|^2\,\dfrac{1}{2 E_{\bf k}}\Big) \,d \xi\,.
\eeq{DiffDecay2} 

\subsection{Decay rate of Gaussian wave packet centered at zero momentum: width correction} 
For a particle of mass $M$ at rest (${\bf k}=0$), we obtain from \reff{DiffDecay} that 
\beq
d\xi= 2\,M d\Gamma_{0_{({\bf k}=0)}}
\eeq{xi}
where $d\Gamma_{0_{({\bf k}=0)}}$ is the plane wave result for the differential decay rate of a particle at rest.  We will now represent a ``particle at rest" by a Gaussian wavepacket centered at ${\bf k}=0$ with width $a$ and normalized according to \reff{Norm}. This is given by
\beq
\phi({\bf k})= \Big(\dfrac{2}{a}\Big)^{3/2} \pi^{3/4} e^{-{\bf k}^2/(2\,a^2)}\,.
\eeq{wavep}
Substituting this into \reff{DiffDecay2} and using \reff{xi} we obtain that
\begin{align}
d\Gamma_{\text{packet}_{({\bf k}=0)}}&= \Big(\int_0^{\infty} \dfrac{2}{a^3 \sqrt{\pi}} \,{\bf k}^2 e^{-{\bf k}^2/a^2}\dfrac{1}{\sqrt{{\bf k}^2+M^2}} d|{\bf k}|\Big) 2M \,d\Gamma_{0_{({\bf k}=0)}}\nonumber\\
&=\dfrac{1}{a}\,\text{U}(1/2, 0, M^2/a^2) M \,d\Gamma_{0_{({\bf k}=0)}}
\label{DiffDecay4} 
\end{align}
where $\text{U}(a,b,z)$ is known as the confluent hypergeometric function of the second kind\cite{Daalhuis}. We are interested in obtaining the correction to the plane wave ($a=0$) result. We therefore perform a series expansion of $\text{U}/a$ about $a=0$ and this yields
\beq
\dfrac{1}{a}\,\text{U}(1/2, 0, M^2/a^2)= \frac {1} {M}\Big(1 - \frac{3\,a^2}{4 M^2} + \mathcal{O}(\tfrac{a^4}{M^4})\Big)\,.
\eeq{Series}
Substituting the above into \reff{DiffDecay4} we obtain our first result: 
\beq
\Gamma_{\text{packet}_{({\bf k}=0)}}= \Gamma_{0_{({\bf k}=0)}}\,\Big(1 - \frac{3 a^2}{4 M^2} + \mathcal{O}(\tfrac{a^4}{M^4})\Big) 
\eeq{finale}
where $\Gamma_{0_{({\bf k}=0)}}$ is the plane wave result for the decay rate of a particle at rest and $\Gamma_{\text{packet}_{({\bf k}=0)}}$ is the decay rate of a Gaussian wave packet centered at zero momentum (note that we dropped the differential symbol because the relation \reff{finale} holds for the total decay rates). We see that the first correction due to the wave packet is of order $\tfrac{a^2}{M^2}$ and vanishes in the plane wave limit where $a$ tends to zero\footnote{We work in units where $\hbar=c=1$. Reinstating these two constants the correction is of order $\tfrac{\hbar^2 a^2}{M^2 c^2}=\lambda_c^2\,a^2$ where $\lambda_c=\tfrac{\hbar}{M c}$ is the Compton wavelength of the particle.}.  Higher-order corrections such as $\tfrac{a^4}{M^4}$, etc. are expected to be much smaller and are not included here. The negative sign in \reff{finale} makes physical sense because we expect the lifetime to increase due to the time dilation produced by non-zero momenta in the wave packet (since there is a spread of momenta about zero). 

\subsection{The Lorentz boost of a wave packet}   

We now determine how $\phi({\bf k})$ transforms under a Lorentz boost. The state of a wavepacket $\ket{\phi}$ is given by \reff{Packet}. After a Lorentz transformation, we label all quantities by a prime i.e.   
\beq
\ket{\phi'}=\int \dfrac{d^3k'}{(2\pi)^3} \dfrac{1}{\sqrt{2E_{\bf k}'}}\phi'({\bf k}')\ket{\bf k'}\,.
\eeq{PacketL}
Quantum states transform under a Lorentz transformation $\Lambda$ via a unitary operator $U(\Lambda)$ \cite{Peskin}. For example, $\ket{\bf k'}$ is given by $U(\Lambda)\ket{\bf k}$. Therefore
\begin{align}
\ket{\phi'}&=U(\Lambda)\ket{\phi}=\int \dfrac{d^3k}{(2\pi)^3} \dfrac{1}{\sqrt{2E_{\bf k}}}\phi({\bf k})\,\,(U(\Lambda)\ket{\bf k})\nonumber\\&=\int \dfrac{d^3k}{(2\pi)^3} \dfrac{1}{\sqrt{2E_{\bf k}}}\phi({\bf k})\,\ket{\bf k'}\,.
\label{PacketL2}
\end{align}
Comparing \reff{PacketL2} with \reff{PacketL} we find that 
\beq
\dfrac{d^3k}{\sqrt{E_{\bf k}}}\,\phi({\bf k})=\dfrac{d^3k'}{\sqrt{E_{\bf k}'}}\,\phi'({\bf k}')\,.
\eeq{PhiPhi}
In other words, the (continuous) coefficients in the expansion of a wave packet are Lorentz invariant. Under a Lorentz boost in the 3-direction we have $k_3'= \gamma (k_3 + v\,E)$ and $E' = \gamma E(1+ v \,k_3/E)$ so that  $\frac{d^3k'}{\sqrt{E_{\bf k}'}}=\frac{d^3k}{\sqrt{E_{\bf k}}}[\gamma\,(1+v \,k_3/E)]^{1/2}$. Substituting this into \reff{PhiPhi} we obtain 
\beq 
\phi'({\bf k}')=\phi({\bf k})\,[\gamma\,(1+v \,k_3/E)]^{-1/2}\,.
\eeq{PhiL}
The above describes how $\phi({\bf k})$ transforms under a Lorentz boost. The wave packet remains normalized, as it should, after the Lorentz transformation i.e. 
\beq
\int \dfrac{d^3k'}{(2\pi)^3}\, |\phi'({\bf k}')|^2=\int \dfrac{d^3k}{(2\pi)^3} \,[\gamma\,(1+v \,k_3/E)]\,\,\dfrac{|\phi({\bf k})|^2}{[\gamma\,(1+v \,k_3/E)]}=\int \dfrac{d^3k}{(2\pi)^3}\, |\phi({\bf k})|^2 =1\,.
\eeq{Norm2}
We now determine $|\phi'({\bf k}')|^2$ in the primed coordinates when $\phi({\bf k})$ is given by the Gaussian \reff{wavep}. The Lorentz boost in the 3-direction yields the following relations: $k_3= \gamma\,(k_3'- v E')$ and $\gamma\,(1+ v \,k_3/E) =[\gamma\,(1- v \,k_3'/E')]^{-1}$. We then obtain
\beq
|\phi'({\bf k}')|^2=\dfrac{8 \pi^{3/2}}{a^3}\,[\gamma\,(1- v \,k_3'/E')]\,  
e^{-(k_1^2 +k_2^2 + [\gamma\,(k_3'- v E')]^2)/a^2}\,.
\eeq{PhiPrime}
This is clearly not a Gaussian. We therefore see that a Gaussian does not transform into another Gaussian under a Lorentz boost. Though the exponential peaks at $k_1=0,k_2=0,k_3'=M\,\gamma \,v$, \reff{PhiPrime} does not represent a Gaussian wave packet centered at momentum $M\,\gamma \,v$. We will study such a Gaussian in section 2.3. We now turn to determining the decay rate of the Lorentz-boosted wavepacket.

\subsubsection{Decay rate of a Lorentz-boosted wave packet and the WP clock}

The differential decay rate of a wave packet that is Lorentz boosted (assumed to be in the 3- direction)  is given by \reff{DiffDecay2} with primes added to represent the Lorentz boosted quantities: 
\beq
d\Gamma_{\text{packet}_{boosted}}= \Big(\int \dfrac{d^3k'}{(2\pi)^3} |\phi'({\bf k'})|^2\,\dfrac{1}{2 E_{\bf k}'}\Big) \,d \xi\,.
\eeq{DiffDecay3}   
where $d\xi$ is given by the Lorentz-invariant quantity \reff{Invariant}. Before boosting, we start with the Gaussian wavepacket centered at zero momentum given by \reff{wavep}
\beq 
|\phi({\bf k})|^2= \dfrac{8 \pi^{3/2}}{a^3}  
\,e^{-(k^2 +k_3^2)/a^2}\,.
\eeq{Phi2}
where $k^2=k_1^2+k_2^2$ so that $d^3k= 2 \pi\,k dk\,dk_3$. From the calculation carried out in \reff{Norm2} we can see that $d^3k' \,|\phi'({\bf k'})|^2= d^3k\, |\phi({\bf k})|^2$. Moreover, $E_{\bf k}'=\sqrt{k^2+ k_3'^2 + M^2}$ where $k_3'=\gamma\,(k_3 +v \,E)$ with $E=\sqrt{k^2 +k_3^2+M^2}$. The decay rate of the Lorentz boosted wave packet is then given by
\beq
d\Gamma_{\text{packet}_{boosted}}= \Big(\dfrac{2}{a^3\,\sqrt{\pi}}\int_0^{\infty}\int_{-\infty}^{\infty}\,dk \,dk_3 \,\dfrac{k\,e^{-(k^2 +k_3^2)/a^2}}{2 \sqrt{k^2+ [\gamma\,(k_3 +v \,E)]^2 + M^2}}\Big) \,d \xi\,.
\eeq{DiffDecay5}  
The integrations converge but do not yield an analytical result in terms of any well-known function. However, since the exponential peaks at $k=k_3=0$ we can perform a series expansion about $k=0$ and $k_3=0$ of the function that multiplies the exponential and then integrate. This leads to an analytical result as a series in powers of the small parameter $a^2/M^2$ up to arbitrary accuracy. The result is (see Appendix A) 
\begin{align}
\Gamma_{\text{packet}_{boosted}}&=\sqrt{1-v^2}\,\dfrac{1}{2M}\Big[1-\dfrac{3a^2}{4M^2}\Big(1-\dfrac{2\,v^2}{3}\Big)+\mathcal{O}(\tfrac{a^4}{M^4})\Big]\,d\xi\nonumber\\
&=\sqrt{1-v^2}\,\,\Gamma_{0_{({\bf k}=0)}}\,\Big[1-\dfrac{3a^2}{4M^2}\Big(1-\dfrac{2\,v^2}{3}\Big)+\mathcal{O}(\tfrac{a^4}{M^4})\Big]\nonumber\\
&=\sqrt{1-v^2}\,\,\Gamma_{\text{packet}_{({\bf k}=0)}}\Bigg[\dfrac{1-\dfrac{3a^2}{4M^2}\Big(1-\dfrac{2\,v^2}{3}\Big)+\mathcal{O}(\tfrac{a^4}{M^4})}{1-\dfrac{3a^2}{4M^2}+ \mathcal{O}(\tfrac{a^4}{M^4})}\Bigg]\nonumber\\
&= \sqrt{1-v^2}\,\Gamma_{\text{packet}_{({\bf k}=0)}}\,\Big(1+\dfrac{a^2\,v^2}{2M^2}+\mathcal{O}(\tfrac{a^4}{M^4})\Big)
\label{Boosted}    
\end{align}
where we used \reff{xi} for $d\xi$ and used \reff{finale} to express $\Gamma_{0_{({\bf k}=0)}}$ in terms of $\Gamma_{\text{packet}_{({\bf k}=0)}}$. In the last equality we performed a binomial expansion and kept terms up to order $a^2/M^2$. 

The important point to take away from the above result is that the decay rate of a wave packet after a Lorentz boost is not related to the original decay rate of the wave packet by \textit{exactly} the Lorentz factor $\sqrt{1-v^2}$. There is a small correction\footnote{Note that compared to the decay rate of a boosted \textit{plane wave} the correction is equal to $-\tfrac{3a^2}{4M^2}(1-\tfrac{2\,v^2}{3})\,.$} equal to $\tfrac{a^2\,v^2}{2\,M^2}\,$. The correction increases as the boost velocity increases (reaching a maximum in the limit $v \to 1$).

We therefore see that the decay of a localized particle in QFT does not act \textit{exactly} like a typical clock under Lorentz boosts and we refer to it as a ``WP clock" (wave packet clock). In contrast to a typical clock, a WP clock does not move with a single velocity but has a spread in velocities (or more specifically, a spread in momenta). It is best to view the WP clock as a single clock since the wave packet represents a one-particle state in QFT. It should be emphasized that there is no breakdown of Lorentz symmetry here. The effect is due to the fact that \textit{localized} particles in QFT must be represented by wave packets and not plane waves in accordance with Heisenberg's uncertainty principle.      

\subsection{Decay rate of Gaussian wave packet centered at velocity $v$}
As we saw, the Lorentz boost by a velocity $v$ of a Gaussian wave packet centered at zero velocity does not yield a Gaussian wave packet centered at velocity $v$. It is therefore of interest to determine the decay rate of a Gaussian wavepacket centered at velocity $v$ and compare it to the decay rate of a Gaussian at ``rest". The third component $k_3$ of the momentum will be centered at $M \gamma v$ where $\gamma$ is the usual Lorentz factor $(1-v^2)^{-1/2}$. The other two components, $k_1$ and $k_2$, will both be centered at zero. The width is denoted by $b$ and the Gaussian normalized again according to \reff{Norm}. This yields
\beq
\phi({\bf k})= \Big(\dfrac{2}{b}\Big)^{3/2} \pi^{3/4} e^{-[{\bf p}^2 +(k_3- M\,\gamma v)^2]/(2\,b^2)}
\eeq{wavep2} 
where ${\bf p}=(k_1,k_2)$ and ${\bf p}^2=k_1^2+k_2^2$.    
Substituting the above into \reff{DiffDecay2} and using $d^3k=dk_1\,dk_2\,dk_3=2\pi |{\bf p}| d|{\bf p}|dk_3$ we obtain 
\begin{align}
d\Gamma_{\text{packet}_{(|{\bf k}|=M\gamma v)}}&= d\xi\,\int_0^{\infty} \int_{-\infty}^{\infty}\dfrac{1}{b^3 \sqrt{\pi}} \,|{\bf p}| \dfrac{e^{-[{\bf p}^2 +(k_3- M\,\gamma v)^2]/b^2}}{\sqrt{{\bf p}^2+k_3^2+M^2}} d|{\bf p}|\,dk_3 \,.
\label{DiffDecay6}
\end{align} 
The above integral converges but does not yield an analytical result in terms of any well-known function. However, as in the previous section, we can obtain an analytical result as a power series expansion in the small parameter $b^2/M^2$ up to arbitrary accuracy. The exponential in this case peaks at $|{\bf p}|=0$ and $k_3=M\,\gamma v$. We can therefore perform a series expansion about $|{\bf p}|=0$ and $k_3=M\,\gamma v$ of the function that multiplies the exponential and then integrate. The result is (see Appendix A for details):
\begin{align}
d\Gamma_{\text{packet}_{(|{\bf k}|=M\gamma v)}}  = d\xi \,\sqrt{1-v^2}\,\,\Big[\dfrac{1}{2M} -\dfrac{3b^2}{8M^3}(1-v^2)^2+\mathcal{O}(\tfrac{b^4}{M^4})\Big]
\label{DiffDecay7}
\end{align} 
Substituting $d\xi$ from \reff{xi} into \reff{DiffDecay7} we obtain
\begin{align}
\Gamma_{\text{packet}_{(|{\bf k}|=M\gamma v)}} = \sqrt{1-v^2}\,\,\Gamma_{0_{({\bf k}=0)}}\Big[1 -\dfrac{3b^2}{4M^2}(1-v^2)^2+\mathcal{O}(\tfrac{b^4}{M^4})\Big]
\label{DiffDecay8}
\end{align} 
where we dropped the differential symbol as before (since the relation applies to the total decay rate). Substituting $\Gamma_{0_{({\bf k}=0)}}$ from \reff{finale} into \reff{DiffDecay8} we obtain 
\begin{align}
\Gamma_{\text{packet}_{(|{\bf k}|=M\gamma v)}} &= \sqrt{1-v^2}\,\,\Gamma_{\text{packet}_{({\bf k}=0)}}\,\Bigg[\dfrac{1 -\frac{3b^2}{4M^2}(1-v^2)^2+\mathcal{O}(\tfrac{b^4}{M^4})}{1 -\frac{3a^2}{4M^2}+\mathcal{O}(\tfrac{a^4}{M^4})}\Bigg]\nonumber\\
&=\sqrt{1-v^2}\,\Gamma_{\text{packet}_{({\bf k}=0)}}\,\Big[1 +\frac{3}{4M^2}\Big(a^2-b^2\,(1-v^2)^2\Big) +\mathcal{O}(\tfrac{b^4}{M^4},\tfrac{a^4}{M^4},\tfrac{a^2\,b^2}{M^4})\Big]\,.
\label{DiffDecay9}
\end{align} 
We therefore see that $\Gamma_{\text{packet}_{(|{\bf k}|=M\,\gamma \,v)}}$ is not \textit{exactly} equal to $\sqrt{1-v^2}\,\,\Gamma_{\text{packet}_{({\bf k}=0)}}$.  In other words, the two decay rates are not related \textit{exactly} by the Lorentz factor $\sqrt{1-v^2}$. For two plane waves (the $a\to 0$ and $b\to 0$ limit) the relation is exact as expected. For the case where the two widths are equal ($a=b$), the correction to the Lorentz factor is non-zero and equal to $\tfrac{3\,a^2\,v^2}{4\,M^2}(2-v^2)$. The correction increases with $v$ and reaches a maximum in the limit $v\to 1$. 

In particular, note that if the width $b$ of the second (moving) Gaussian is equal to $a\,\gamma^2$ where $a$ is the width of the Gaussian at ``rest", then the leading order correction vanishes and the two decay rates are related by the Lorentz factor (up to higher order corrections). So the decay rate of a Gaussian centered at velocity $v$ and one centered at zero velocity are not in general related exactly by the Lorentz factor except when the ratio of their widths $b/a$ is equal to $\gamma^2$. 

\section{Conclusion}
In this paper, we found an analytical result for the leading order width correction to the decay rate of a Gaussian wave packet centered at zero momentum. The effect was of order $\tfrac{a^2}{M^2}$ where $a$ is the width and $M$ is the mass of the decaying particle. We then showed that the decay rate of this Gaussian wave packet under a Lorentz boost by velocity $v$ does not decrease exactly by the Lorentz factor $\sqrt{1-v^2}$. There is a correction of order $\tfrac{a^2v^2}{M^2}$. This is important because it means that a decaying wave packet does not behave like a typical clock under Lorentz boosts and we refer it to it as a ``WP clock". A WP clock can be defined as a clock that does not move with a single velocity relative to an observer but has a spread in velocities (or a spread in momenta). Since the wave packet is a one-particle state in QFT it is best to view the WP clock as a single clock. As already mentioned, the WP clock does not imply a breakdown of Lorentz symmetry: it does not stem from new physics but is an important consequence of the combined requirements of special relativity, quantum mechanics and \textit{localized} free particles. 

We found that a Gaussian wave packet which is Lorentz-boosted does not transform into another Gaussian. This is why we calculate separately the decay rate of a Gaussian centered at momentum $M\,\gamma\,v$ (with width $b$). We find that it is not exactly equal to the decay rate of a Gaussian centered at zero momentum 
(with width $a$) multiplied by the Lorentz factor $\sqrt{1-v^2}$. There is again a correction that depends on their respective widths and the velocity $v$. For the case where their respective widths are equal ($a=b$), the leading order correction to the Lorentz factor is equal to $\tfrac{3\,a^2\,v^2}{4\,M^2}(2-v^2)$. However, the leading order correction vanishes if the ratio of their widths $b/a$ is equal to $\gamma^2$.    

There is one thing in particular that would be important to investigate in future work. In special relativity, there is a symmetry between two inertial frames. If two frames A and B move with constant velocity relative to each other, then from A's perspective, B's clock runs slower by the usual time dilation factor and from B's perspective,  A's clock runs slower by the same time dilation factor. Consider now a Gaussian wave packet that is centered at zero momentum relative to an inertial frame A. Let its decay rate in A's frame be $\Gamma_A$. If we now Lorentz boost the wave packet with velocity $v$ the decay rate of the Lorentz-boosted wave packet will be $\Gamma_A$ multiplied by the usual Lorentz factor of $\sqrt{1-v^2}$ but \textit{with a correction} due to the width and speed $v$. In other words, from A's perspective, the clock now runs slower but with a small correction to the usual time dilation: this correction is what defines a WP clock. Now, frame A, by definition, is an inertial frame with a typical clock. By symmetry, its clock relative to a frame B attached to the free particle wave packet should also run slower by the exact same amount as the WP clock. Since A's clock is a typical clock, this implies that the correction that appears in a WP clock must now originate from the nature of frame B. In other words, if we want to preserve the symmetry inherent to special relativity, frame B attached to the free particle wave packet must be a slightly different kind of inertial frame (just like a WP clock is a slightly different kind of clock). One argument in favor of regarding frame B as an inertial frame is that it is a frame attached to a \textit{free} particle. However, in contrast to a free particle in classical mechanics, a free localized particle in QFT does not have a precise velocity/momentum. So if frame B can be regarded as an inertial frame, it makes sense that it would be a slightly different kind of inertial frame. The goal would then be to see if one can define such an inertial frame so that it is consistent with special relativity and leads to the desired effects. This is a worthy goal to pursue with the potential of broadening our scope.

\begin{appendices}
\numberwithin{equation}{section}
\setcounter{equation}{0}

\section{Integrations via series expansions}
%\numberwithin{equation}{section}
%\setcounter{equation}{0}
\subsection{Integrals appearing in the decay rate of a boosted wave packet}
In section 2.2.1 we encountered the decay rate of a boosted wave packet given by the double integral \reff{DiffDecay5}:
\begin{align}
&d\Gamma_{\text{packet}_{boosted}}= \Big(\dfrac{2}{a^3\,\sqrt{\pi}}\int_0^{\infty}\int_{-\infty}^{\infty}\,dk \,dk_3 \,\dfrac{k\,e^{-(k^2 +k_3^2)/a^2}}{2 \sqrt{k^2+ [\gamma\,(k_3 +v \,E)]^2 + M^2}}\Big) \,d \xi\,\nonumber\\
&=\Big(\dfrac{1}{2\,a^3\,\sqrt{\pi}}\int_0^{\infty}\int_{-\infty}^{\infty}\,du \,dk_3 \,\dfrac{e^{-(u +k_3^2)/a^2}}{ \sqrt{u+ [(1-v^2)^{-1/2}\,(k_3 +v \sqrt{u+k_3^2+M^2})]^2 + M^2}}\Big) \,d \xi \label{DD}
\end{align}  
where we made the substitution $u=k^2$, replaced $E$ by its expression $\sqrt{k^2+k_3^2+M^2}=\sqrt{u+k_3^2+M^2}$ and replaced $\gamma$ by $(1-v^2)^{-1/2}$. The integrals converge but do not yield an analytical result in terms of any well-known function. However, by expanding the function that multiplies the exponential about $u=k_3=0$ (where the exponential peaks) as a power series before integrating we can obtain an analytical result for the integral as a power series in the small parameter $a^2/M^2$ up to arbitrary accuracy. The function is (including the normalization pre-factor)  
\beq 
f(u,k_3)=\dfrac{1}{2\,a^3\,\sqrt{\pi}}\Big[u+ [(1-v^2)^{-1/2}\,(k_3 +v \sqrt{u+k_3^2+M^2})]^2 + M^2\Big]^{-1/2}\,.
\eeq{Fk}
Our goal is to obtain the leading correction to the plane wave result for the decay rate. This means we want the integral (with the normalization pre-factor) to be accurate up to $a^2/M^2$ (but do not require it to be accurate up to $a^4/M^4$ or higher). This requires that we expand the function $f$ above only up to second order in $k_3$ and first order in $u$. This yields the function $f_2$ given by 
\begin{align}
f_2=\dfrac{\sqrt{1 -v^2}}{\sqrt{\pi}\,a^3}\, \dfrac{1}{2\,M}\Big[1 - \dfrac{u}{2\,M^2} +  &- \dfrac{k_3}{M^3}(M^2 - u ) v \nonumber\\&\quad\quad+
    \dfrac{k_3^2}{4\,M^4} (2 M^2 - 3 \,u) (-1 + 
      2 v^2)\Big]\,.
\label{ff}
\end{align}
Multiplying $f_2$ by the exponential and then performing the integration yields 
\beq
\int_0^{\infty}\int_{-\infty}^{\infty}\,du \,dk_3 \,f_2\,e^{-(u +k_3^2)/a^2}=
\sqrt{1-v^2}\,\dfrac{1}{2M}\Big(1-\dfrac{3a^2}{4M^2}(1-2\,v^2/3)+\mathcal{O}(\tfrac{a^4}{M^4})\Big)\,.
\eeq{DD2}
The differential decay rate of a boosted wave packet given by \reff{DD} is obtained by multiplying the above result by the factor $d \xi=2\,M d\Gamma_{0_{({\bf k}=0)}}$ given by \reff{xi} where $d\Gamma_{0_{({\bf k}=0)}}$ is the differential decay rate of a particle at rest treated as a plane wave. This yields 
\beq  
\Gamma_{\text{packet}_{boosted}}=\sqrt{1-v^2}\,\Gamma_{0_{({\bf k}=0)}}\Big(1-\dfrac{3a^2}{4M^2}(1-2\,v^2/3)+\mathcal{O}(\tfrac{a^4}{M^4})\Big)
\eeq{deca}
where we have dropped the differential sign since the relation applies also to the total decay rate. We would like to express the decay rate of the boosted wave packet in terms of the decay rate of the original wave packet ``at rest" (centered at zero momentum). Using \reff{finale} we can express the decay rate of a plane wave at rest to the decay rate of a wave packet ``at rest" i.e.
\beq
\Gamma_{0_{({\bf k}=0)}}= \dfrac{\Gamma_{\text{packet}_{({\bf k}=0)}}}{1 - \frac{3 a^2}{4 M^2} + \mathcal{O}(\tfrac{a^4}{M^4})}\,.
\eeq{finale2}
Substituting the above into \reff{deca} we obtain 
\begin{align}
\Gamma_{\text{packet}_{boosted}}&=\sqrt{1-v^2}\,\,\Gamma_{\text{packet}_{({\bf k}=0)}}\Bigg[\dfrac{1-\dfrac{3a^2}{4M^2}(1-2\,v^2/3)+\mathcal{O}(\tfrac{a^4}{M^4})}{1-\dfrac{3a^2}{4M^2}+ \mathcal{O}(\tfrac{a^4}{M^4})}\Bigg]\nonumber\\
&=\sqrt{1-v^2}\,\Gamma_{\text{packet}_{({\bf k}=0)}}\,\Big(1+\dfrac{a^2\,v^2}{2M^2}+\mathcal{O}(\tfrac{a^4}{M^4})\Big)\,\label{deca2}
\end{align}
where in the last line we performed a binomial expansion and kept terms only up to the small parameter $a^2/M^2$. We therefore see that the Lorentz boost of a wave packet does not change its decay rate exactly by the Lorentz factor $\sqrt{1-v^2}$. There is a correction of order $a^2\,v^2/M^2$. 

\subsection{Integrals appearing in the decay rate of a Gaussian wave packet centered at velocity v}
In section 2.3 we encountered the decay rate of a Gaussian wave packet centered at velocity v  given by the  double integral \reff{DiffDecay6}
\begin{align}
d\Gamma_{\text{packet}_{(|{\bf k}|=M\gamma v)}}&= d\xi\,\int_0^{\infty} \int_{-\infty}^{\infty}\dfrac{1}{b^3 \sqrt{\pi}} \,|{\bf p}| \dfrac{e^{-[{\bf p}^2 +(k_3- M\,\gamma v)^2]/b^2}}{\sqrt{{\bf p}^2+k_3^2+M^2}} d|{\bf p}|\,dk_3 \nonumber\\
&=d\xi\,\int_0^{\infty} \int_{-\infty}^{\infty}\dfrac{1}{2 b^3 \sqrt{\pi}} \dfrac{e^{-[u +p_3^2]/b^2}}{\sqrt{u+(p_3+M\,\gamma v)^2+M^2}} du\,dp_3
\label{DiffDecayA}
\end{align} 
where we made the substitution $u=|{\bf p}|^2$ and $p_3=k_3-M\,\gamma v$. The integrals converge but do not yield an analytical result in terms of any well-known function. Our procedure will be the same as the one used for the previous integral in the subsection above. The exponential peaks at $u=p_3=0$. We therefore expand the function that multiplies the exponential as a power series about $u=0$ and $p_3=0$. We can then obtain an analytical result for the integral as a power series in the small parameter $b^2/M^2$ up to arbitrary accuracy. The function in this case is 
\beq 
f= \dfrac{1}{2 b^3 \sqrt{\pi}}\Big[u+(p_3+M\,\gamma v)^2+M^2\Big]^{-1/2}\,.
\eeq{ff3}
We are interested in obtaining only the leading correction to the decay rate proportional to $b^2/M^2$. This requires us to expand $f$ only up to second order in $p_3$ and first order in $u$. We label this $f_3$ and it is given by
\begin{align}
f_3=\dfrac{1}{8 \,b^3 \,M^5 \,\sqrt{\pi}}&\Big(4 \,M^4 \,\sqrt{1 - v^2} 
  -4 \,M^3 \,p_3 \,v \,(1 -v^2) + 6 \,M \,p_3 \,u \,v (1 - v^2)^2\nonumber\\ &+
  3 \,p_3^2\, u (1 - v^2)^{5/2} (1 - 5 v^2) - 
  2 \,M^2 (1 - v^2)^{3/2} (u + p_3^2 (1 - 3 v^2))\Big)\,.
	\label{GaussExp}
	\end{align}
The decay rate \reff{DiffDecayA} is then given by 
\begin{align}
\Gamma_{\text{packet}_{(|{\bf k}|=M\gamma v)}}&=d\xi\,\int_0^{\infty}\int_{-\infty}^{\infty}\,du \,dp_3 \,f_3\,e^{-(u +p_3^2)/b^2}\nonumber\\
&=\sqrt{1-v^2}\,\dfrac{d\xi}{2M}\Big(1-\dfrac{3b^2}{4M^2}(1-v^2)^2+\mathcal{O}(\tfrac{b^4}{M^4})\Big)
\label{DecayA2}
\end{align}
where we dropped the differential sign for the decay rate as the result holds for the total decay rate. Using \reff{xi} and \reff{finale} we now replace $d\xi$ by
\beq
d\xi=2M\,\Gamma_{0_{({\bf k}=0)}}= 2M\,\dfrac{\Gamma_{\text{packet}_{({\bf k}=0)}}}{1 - \frac{3 a^2}{4 M^2} + \mathcal{O}(\tfrac{a^4}{M^4})}\,.
\eeq{xi3}
Substituting the above into the decay rate \reff{DecayA2} we obtain our final result
\begin{align}
\Gamma_{\text{packet}_{(|{\bf k}|=M\gamma v)}} &= \sqrt{1-v^2}\,\,\Gamma_{\text{packet}_{({\bf k}=0)}}\,\Bigg[\dfrac{1 -\frac{3b^2}{4M^2}(1-v^2)^2+\mathcal{O}(\tfrac{b^4}{M^4})}{1 -\frac{3a^2}{4M^2}+\mathcal{O}(\tfrac{a^4}{M^4})}\Bigg]\nonumber \\
&=\sqrt{1-v^2}\,\Gamma_{\text{packet}_{({\bf k}=0)}}\,\Big[1 +\frac{3}{4M^2}\Big(a^2-b^2\,(1-v^2)^2\Big)+\mathcal{O}(\tfrac{b^4}{M^4},\tfrac{a^4}{M^4})\Big]\,.
\label{DecayA3}
\end{align} 

\end{appendices}

%---------------------------------------------------------------------------------------
\section*{Acknowledgments}
The author acknowledges support from a discovery grant of the National Science and Engineering Research Council of Canada (NSERC). The author also thanks Michael E. Peskin for valuable communications.

\end{document}